# NeoRS: a neonatal resting state fMRI data preprocessing pipeline

V. Enguix[1,2,3], J. Kenley[4], D. Luck,[1,3] J. Cohen-Adad[2,5,6], G.A. Lodygensky[1,3].

1. Department of Pediatrics, CHU Sainte-Justine, University of Montreal, Canada
2. NeuroPoly Lab, Institute of Biomedical Engineering, Polytechnique Montreal, Canada
3. Canadian Neonatal Brain Platform, Canada
4. Washington University School of Medicine, Saint Louis, USA
5. Functional Neuroimaging Unit, CRIUGM, University of Montreal, Montreal, QC, Canada
6. Mila – Quebec AI Institute, Montreal, QC, Canada

## Abstract

Resting state functional MRI (rsfMRI) has been shown to be a promising tool to study intrinsic brain functional connectivity and assess its integrity in cerebral development. In neonates, where functional MRI is limited to very few paradigms, rsfMRI was shown to be a relevant tool to explore regional interactions of brain networks. However, to identify the resting state networks, data needs to be carefully processed to reduce artifacts compromising the interpretation of results. Because of the non-collaborative nature of the neonates, the differences in brain size and the reversed contrast compared to adults due to myelination, neonates can't be processed with the existing adult pipelines, as they are not adapted. Therefore, we developed NeoRS, a rsfMRI pipeline for neonates. The pipeline relies on popular neuroimaging tools (FSL, AFNI, SPM) and is optimized for the neonatal brain. The main processing steps include image registration to an atlas, skull stripping, tissue segmentation, slice timing and head motion correction and regression of confounds which compromise functional data interpretation. To address the specificity of neonatal brain imaging, particular attention was given to registration including neonatal atlas type and parameters, such as brain size variations, and contrast differences compared to adults. Furthermore, head motion was scrutinized, and motion management optimized, as it is a major issue when processing neonatal rsfMRI data. The pipeline includes quality control using visual assessment checkpoints. To assess the effectiveness of NeoRS processing steps we used the neonatal data from the Baby Connectome Project dataset including a total of 10 neonates. NeoRS was designed to work on both multi-band and single-band acquisitions and is applicable on smaller datasets. NeoRS also includes popular functional connectivity analysis features such as seed-to-seed or seed-to-voxel correlations. Language, default mode, dorsal attention, visual, ventral attention, motor and fronto-parietal networks were evaluated. Topology found the different analyzed networks were in agreement with previously published studies in the neonate. NeoRS is coded in Matlab and allows parallel computing to reduce computational times; it is open-source and available on GitHub [https://github.com/venguix/NeoRS]. NeoRS allows robust image processing of the neonatal rsfMRI data that can be readily customized to different datasets.

# 1. Introduction

The analysis of resting-state functional connectivity (RS-FC) constitutes a promising tool as it provides complementary information to structural imaging related to brain physiology. Indeed, since its discovery in 1995 [1] rsfMRI studies have provided new insights in the understanding of brain architecture and cerebral development [2-7] . Smyser et. al demonstrated the feasibility of using rsfMRI to explore the alterations in resting state networks (RSN) associated with preterm birth and white matter injury [8]. Alterations of the default mode and ventral attention networks at birth, are associated with behavioral inhibition at age of two years, [9] which suggests early alterations of the RSN present a correlation with clinical manifestations, and opens the opportunity of early diagnostics and treatment. Additionally, neonatal RSN are consistently identifiable and present with high similarities to older populations [10-12]. RS-FC is based on low frequency regional fluctuations (<0.1 Hz) in the Blood-Oxygen-Level-Dependent (BOLD) [13, 14] signal while the participant is not performing any task, a useful feature when evaluating neonates [6]. RSN signal is very stable across subjects [15], but vulnerable to several artifacts such as head-motion [16], susceptibility distortions and or white matter (WM) and cerebrospinal fluid (CSF) signals [17, 18]. Robust rsfMRI data processing is key to reduce the nuisance effects of the non-neural signals in the data to identify reliable resting state activity [19, 20]. Its clinical potential and implementation present several methodological challenges that need to be addressed before considering its use to develop a new generation of biomarkers. For this reason, straightforward to use and open-source tools for the neonatal rsfMRI data processing need to be readily available. Tools for mature brains already exist to process rsfMRI data, but analyzing the neonatal brain presents challenges that need to be addressed with altered approaches[6]. There are several straightforward rsfMRI data processing pipelines developed for adults such as Conn toolbox [21],fmriprep, [22] or HCP [23], however, those are not adapted to the newborn brain which presents additional challenges, such as different contrast due to myelination [24]. T2-weighted images are usually needed for tissue segmentation in place of T1-weighted images. Further, varying brain sizes between subjects [6] makes adult skull stripping less robust on the neonatal brain. Additionally, different age specific atlases and tissue probability maps are required for accurate segmentations, common space normalization and seed-based analysis.

To the best of our knowledge the only existing open-access pipeline to process neonatal rsfMRI data is the one developed by the developing Human Connectome Project (dHCP) [25]. While this pipeline has proven to provide excellent results with the dHCP data, its implementation on smaller or clinical datasets remains challenging, as it requires large datasets for independent component (IC) denoising. Furthermore, the dHCP pipeline can be difficult to set up for cohorts acquired at other centers, because the pipeline was developed/optimized from the dHCP database specifically. For example, the dHCP denoising step is based on spatial independent component analysis (sICA), which separates independent correlating signals that can be classified as neural or non-neural signal. This denoising technique has been shown to provide superior results in adults and infants when the dimensionality is accurately set [26, 27]. However, the identified signals need to be classified as neural signal or structured noise, which in most cases is performed manually and a difficult process to automate. To overcome this limitation, the dHCP pipeline uses a machine learning approach (ICA-based Xnoiseifier) [28] to classify the independent components as neural signals or noise. The machine learning algorithm requires a minimum of 35 manually labelled subjects to be trained, which is not always possible in smaller cohorts and requires specialists to manually classify the independent components [25].

To overcome the aforementioned challenges, we developed NeoRS, with the goal of creating a robust open-source pipeline containing the necessary tools to preprocess rsfMRI data. The main advantages of NeoRS are it has been developed specifically for neonates, simple to implement and flexible to process different datasets. Additionally, it can process single subject data, utilizes parallelizable environment and includes visual quality control checkpoints at each step.

The data processing steps include T2-weighted image alignment to a common space, slice timing correction and segmentation, and rsfMRI procedures are slice timing, distortion correction using reversed phase encoding polarity acquisitions, alignment in a common space, motion correction, removal of nuisance confounds and noise compromising functional data interpretation. Further, simple resting state functional connectivity cross-correlations based on seed-to-voxel and seed-to-seed approaches are incorporated.

## 2. Materials and Methods

### Data

NeoRS has been evaluated on neonates (7+/-1.4 weeks old) from the *Baby Connectome Project (BCP)* [29] dataset. For this study only participants scanned at 9 weeks old or less and contained T2-weighted images and rsfMRI were used (N=10). Participants were naturally sleeping and were scanned on a 3.0 T MRI Prisma from Siemens using a 32-channel head coil. This study included a T2-weighted structural image (TE = 564 ms, TR = 3200 ms, matrix=320x320 mm, FOV = 256x256 mm, resolution = 0.8 x 0.8 x 0.8 mm, flip angle = variable, in-plane acceleration factor=2, acquisition time = 5 min 57 s), two gradient-echo (GRE) echo-planar imaging (EPI) blip-up/blip-down (TE =37 ms, TR=800 ms, matrix=104x91 mm, FOV = 208x182 mm, resolution = 2x2x2 mm, flip angle = 52°, multiband acceleration factor=8, acquisition time = 5 min 47 s, 420 volumes), and two spin-echo (SE) EPI blip-up/blip-down for distortion correction purpose (TE =66 ms, TR=8000 ms, matrix=104x91 mm, FOV = 208x182 mm, resolution = 2x2x2 mm, flip angle = 52°, multiband acceleration factor=1, acquisition time = 33 s, 3 volumes).

### Data structure

To facilitate collaborations, NeoRS uses the Brain Imaging Data Structure (BIDS) format as described in https://bids.neuroimaging.io/. See [Figure 1] for an example of data naming and organization for NeoRS.

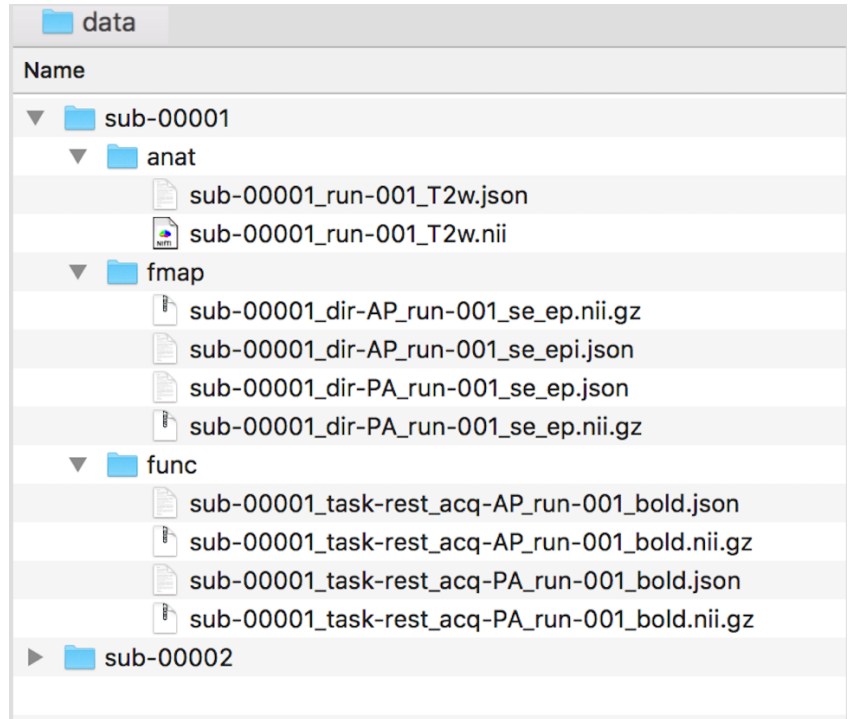

**Figure 1.** Example of data naming and organization for NeoRS.

## Pipeline overview

NeoRS is a neonatal rsfMRI data processing pipeline developed on Matlab and calls for commands developed on well-known open-source neuroimaging tools, such as FSL 6.0.3.1 [30-32], AFNI 20.2.10 [33] and SPM 12 (https://www.fil.ion.ucl.ac.uk/spm/). It runs on both MacOS and Linux operating systems. NeoRS has been tested on a MacBook pro 2015 with operating system High Sierra and on a Linux computer using Ubuntu 18.04.5 by running the pipeline from the beginning to end on different subjects on both computers. NeoRS is built to accommodate MRI data acquired with different manufacturers. The pipeline was tested using the aforementioned BCP data, as well as the not publicly available data from CHU Sainte-Justine acquired on a GE 3T MR750, but this manuscript focuses only on BCP results. Furthermore, the pipeline has single subject capabilities. NeoRS has been developed to accommodate a parallelizable environment, allowing several subjects to be simultaneously processed depending on the number of selected cores. To investigate, two subjects were processed on an early 2015 MacBook pro with 2.7 GHz Intel Core i5 processor and 8 GB 1867 MHz DDR3 memory by using a single core vs 2 parallel cores and found a reduction of computing time of 1.8 times when using the 2 parallel cores. This function is optional and requires the Matlab parallel toolbox.

See NeoRS workflow in [Figure 2].

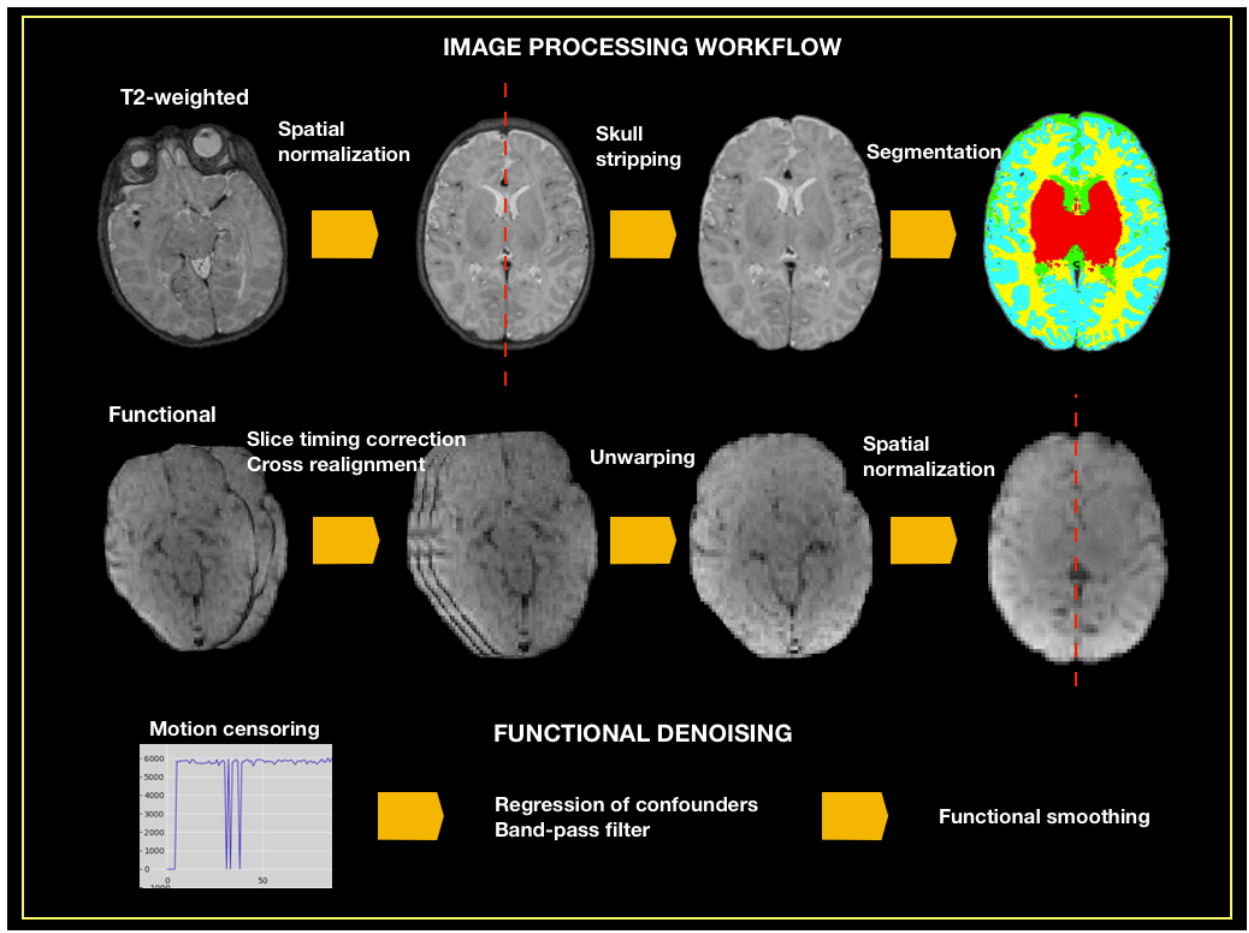

**Figure 2.** NeoRS workflow for neonatal resting state functional connectivity processing and denoising.

To ensure images align to the orientation of the standard template a reorientation to standard is performed in both structural and functional data prior to other data processing procedures by employing fslreorient2std from FSL. Furthermore, to guarantee the accurate performance of NeoRS, output files for each processing step are saved in a folder called Output_files. Processing steps are dependent from previous outputs and should be inspected carefully. In case of fail, the user can parameterize the specific function, as specified later on. Various operations are not mandatory, such as slice-timing correction or distortion correction, and can be manually turned-off by setting the function parameter to 0 in the main file. See [Figure 3] for an example of inputs configuration.

```
%% INPUTS %%%%%%%%%%%%%%%%%%%%%%%%%%%%%%%%%%%%%%%%%%%%%%%%%%%%%%%%%%%%%
%%%%%%%%%%%%%%%%%%%%%%%%%%%%%%%%%%%%%%%%%%%%%%%%%%%%%%%%%%%%%%%%%%%%%%%
% Activate or deactivate functions: 1=On; 0=Off
options.slicetimingcorrection=1; %Slice timing correction
options.fmap = 0; %Functional distortion correction
options.FDaverage = 1;

%Inputs definition
workingDir=('/Desktop/Data');%Data directory
options.TR=3;%Repetition time of the RS sequence in seconds
options.motion=12; %Number of motion parameters-> 6,12 or 24
options.slice_order=5;%1: bottom up, 2: top down, 3: interleaved+bottom up
% 4: interleaved+top down, 5:automatically read json file
options.FWHM=6; %FWMH for functional gaussian smoothing
options.radius=35; %Head radius
options.FD_max=0.25; % Framewise displacement threshold
options.BPF=[0.01,0.1]; %Band-pass filter frequencies in Hz
options.n_core=2; %Number of cores for parallel computing
```

**Figure 3.** Example of NeoRS inputs.

## Data Processing

## Structural

### T2-weighted Image Registration

NeoRS uses the term age stereotaxic space [34] from Washington University – School of Medicine. The template is available in Talairach space [35] 1mm and 3mm isotropic resolutions. Image registration in NeoRS is performed using FSL *flirt* and is implemented in a single step with 12 degrees of freedom and *not* applying the resampling blur when down sampling. These parameters can be modified by the user in the function *anat2std.m*. High resolution T2-weighted images are registered to a 1mm and 3 mm isotropic template.

### Skull Stripping

Skull stripping plays an important role in image processing, as it is mandatory for different processing functionalities, such as tissue segmentation, and requires special attention to avoid further complications in the process. NeoRS skull stripping step utilizes the FSL [32] function *bet2* and has been optimized for term neonatal brains. Skull stripping is performed after image registration to obtain consistent results independent of brain size. Furthermore, visual quality control is available in a file containing the brain with the skull and the overlay of the contour of the intracranial cavity. If the user is not satisfied with the results, modify the fractional intensity threshold, "-f", and vertical gradient in fractional intensity threshold, "-g", to properly adjust skull stripping in the Matlab function *skull_stripping.m*.

## Segmentation

Extracted T2-weighted intracranial content is then segmented to create different tissue probability maps corresponding to each brain structure. Tissue segmentation is crucial in image processing as the outputs will be used for regression of confounds. For brain segmentation NeoRS applies *Morphologically Adaptive Neonatal Tissue Segmentation: Mantis* [36]. *Mantis* is an SPM based toolbox and allows T2-weighted image segmentation based on template adaptation via topological filters and morphological segmentation tools, resulting in eight different tissue probability maps. The segmentation process is fully integrated in the NeoRS pipeline and has been tested on three different datasets (BCP, and CHU Sainte-Justine). After segmentation, the eight tissue probability maps are automatically combined, thresholded and binarized to create three different binary masks needed to run downstream processing. The masks correspond to white matter (WM), grey matter (GM) and cerebrospinal fluid (CSF). The masks are resampled to 3-mm isotropic to match functional image space resolution.

## Functional

### Slice Timing Correction

A common acquisition technique for rsfMRI is the single-shot Gradient-Echo (GRE) Echo-Planar Imaging (EPI). In this acquisition sequence, slices are acquired at varying intervals, which need to be addressed. The NeoRS function for slice timing correction is FSL *slicetimer* and can automatically read the slice order from the .json file, if available. If the .json file is not available, the user can manually define the slice order or use one of the predefined options from fsl (i.e: interleaved ascending) in the configuration file.

## Functional Cross Realignment

To correct for head movement, it is necessary to obtain a motion estimation based on 6 movement parameters (three rotation and 3 translation parameters). This is done by rigid-body registration (6 degrees of freedom) between the different volumes with respect to a reference, in NeoRS, the reference is the first volume from the rsfMRI, but can be easily altered by the user if desired. This NeoRS function is performed using FSL *mcflirt* [37] and works the same as for adults, however we set the smoothness level to 0, as smoothing occurs later in the pipeline, and used sinc interpolation. Parameters for cross realignment can be customized in the Matlab function *cross_realign2.m*. For quality control purposes, NeoRS creates a .png file where the total rotations, translations and framewise displacement (FD) for each volume can be evaluated. Framewise displacement is calculated as previously described by Power et al.[38]. To take into account head size differences, the calculations were done using a 35 mm radius sphere instead of 50 mm which approximately corresponds to the mean distance from the cerebral cortex to the center of the head in neonates. After motion correction, the motion parameters are saved in a text file that will be further used for denoising purposes.

## Functional Best Resting State Section Selection

NeoRS incorporates the possibility to analyze sub sections of long time series (i.e 20 minutes). This tool is deactivated by default but can be activated by setting options.best_volumes = 1 in the configuration file. The sectional analysis tool automatically identifies a section of the time-series (i.e 5 minutes) with the lowest average FD, which is recommended to use on very long acquisitions that present a higher average FD than the threshold. The length of the section can be modified by the user, but it is recommended the duration remain above five minutes [9]. To note, the average FD threshold has a default of 0.25mm, but can also be tailored as needed by altering options.FDaverage in the configuration file. NeoRS chooses only the best sections of the time-series, which reduces computational times drastically.

## Functional Distortion Correction

The EPI sequence is considerably sensitive to off-resonance fields due to susceptibility variations of participants. To address these distortions a typical approach is to use two SE-EPI reversed polarity acquisitions (reversed phase encoding direction) to estimate the distortion field. This field is implemented to correct for distortions in the original GRE EPI images. Directly using two reversed polarity GRE EPI to estimate the distortion field instead of two reversed polarity SE EPI is possible, but not recommended because GRE-EPI sequences are hampered by signal dropouts caused by intravoxel dephasing. Providing reversed phase encoding polarity SE EPI images and activating the distortion correction option (options.fmap=1) in NeoRS, allows users to estimate these distortions utilizing FSL *topup* [31, 39, 40]. Such as specified in the *topup* documentation, a text file containing the encoding directions and total readout time needs to be included in the fmap folder to perform distortion corrections. If the .json file is found in the fmap folder, the text file will be automatically created by NeoRS. Distortion estimates are rectified with FSL *applytopup* for each EPI volume by applying the output from *topup*.

**Functional Image Registration**

Functional images are registered to the same stereotaxic space template from the Washington University – School of Medicine as the T2-weighted image registration procedure. Initially, a two-step registration was performed. First, a rigid body registration between the mean rsfMRI volume and T2-weighted images was calculated, followed by affine transformation between T2-weighted and the template. Finally, the output affine transformation matrices from each process were applied to the rsfMRI images to align to the 3 mm isotropic template. Additionally, a single step registration approach was investigated, where the rsfMRI images were aligned directly to the 3 mm isotropic template using 12 degrees of freedom. This registration approach was comparable to the 2-step registration process and was chosen as it was accurate and faster. Down-sampling blur, by default is set to off in NeoRS, however, if needed, the parameters can be customized in the Matlab function *epi2std2.m*.

**Denoising**

**Motion Censoring**

Before the regression of the confounding signals, volumes with excessive motion are removed based on the framewise displacement metric described by Power et al [38]. NeoRS performs linear detrending and computes framewise displacement after functional cross realignment based on 6 motion parameters in radians (3 rotation parameters + 3 translation parameters). This step also automatically removes the first five volumes.

$$FD = |rot\_x| + |rot\_y| + |rot\_z| + |trans\_x| + |trans\_y| + |trans\_z|$$

Where rot_x/rot_y/rot_z are rotations converted from radians to mm and trans_x/trans_y/trans_z are translations in mm. Once the FD is computed, a text file containing the information of volumes exceeding the FD threshold plus the first 5 frames, is created and head motion plots are saved and can be reviewed. The FD threshold is automatically set to FD < 0.25 mm [41], so volumes with FD higher than or equal to 0.25 mm are excluded. Excluded volumes are set to zero value and no interpolation is applied to avoid artificial correlations.

## High Motion Subjects

High motion acquisitions are source of artifacts and may confound neural correlations with non-neural signals. Apart from single frame motion censoring based on FD NeoRS evaluates the average FD for every BOLD run. By default, NeoRS was set to discard acquisitions with an average FD higher than 0.25 mm. The average FD threshold can be altered in the configuration file by defining options.FDaverage.

## Regression of Confounds

Variables identified as potential confounders of the estimated BOLD signal are merged in a single file. To avoid frequency mismatch in the regression process, the file is used to compute linear regression in a single step with frequency filtering. This process is performed utilizing AFNI *3dTproject*.

### Motion Parameters

Head motion is considered as a rigid body moving in a 3D space with 6 degrees of freedom. In cartesian coordinates we can describe it with 3 translations x- (left/right), y- (anterior/posterior), and z-axes (inferior/superior), and 3 rotations around the x-axis (pitch), y-axis (yaw), and z-axis (roll). To address the residual motion related signal variance after a suboptimal rigid body registration, NeoRS uses a linear regression strategy based on the 6 aforementioned estimated motion parameters. Those parameters are considered as nuisance effects of the signal and are then removed. NeoRS allows various options including: 6 motion parameters, 12 (including temporal derivatives)[38] or 24 (including temporal derivatives and their squares) [42] which can be defined in the configuration file parameter options.motion.

**White Matter, Cerebrospinal Fluid and Global Signals**

White matter and cerebrospinal fluid signals are highly confounding and need to be removed from the rsfMRI [41]. Signals for regression of confounds are extracted from the WM and CSF masks generated previously in the pipeline from the segmentation. Masks are created in a conservative way by selecting voxels from the tissue probability masks with higher probability than 0.5. The voxels of the white matter are eroded by one voxel to ensure the mask doesn't include any gray matter. Two files containing the average signal of the WM and CSF masks are created. Finally, global signal is approximated by averaging the signal in a gray matter mask [43] and used by default in NeoRS, as it improves data quality by reducing motion artifacts (cardiac, respiration, head motion) [41].

**Frequency Filter**

Temporal frequencies outside the frequency range of [0.01 - 0.1] Hz are removed from the BOLD signal to correct for slow frequency drifts, reduce motion artifacts and other physiological noises while preserving the frequencies of resting state networks [18]. The use of a low-pass filter could drastically reduce the degrees of freedom of the time series in acquisitions with very short TR. For those cases, it is recommend to set the value of options.BPF=[HPH, LPF], to [0.01, 999] in the configuration file.

# Smoothing

## Functional Smoothing

Functional smoothing is the last processing procedure. After denoising the rsfMRI signal is convolved with a gaussian kernel. This reduces the effect of misregistration between functional regions and slightly increases the signal to noise ratio. Gaussian smoothing is performed implementing *fslmaths* from FSL. The size of the gaussian kernel is customizable in the NeoRS pipeline by modifying options.fwhm in the configuration file, which is 6 mm by default.

## Data Analysis - Functional Connectivity

Prior to further data analysis, like ROI to ROI (region of interest) correlations, all the processed BOLD runs are merged together into a single 4D-file. NeoRS offers basic single subject data analysis, including seed based and seed to seed correlations, so the user can further assess data has been correctly processed.

### Seed-Based Correlations (SBC)

Seed-based functional connectivity identifies correlation between a defined ROI, also called a seed, and the rest of the brain. This metric facilitates the observation of simultaneously activated regions with the pre-defined ROI. The NeoRS pipeline provides 31 template seeds representing some of the most common resting state networks including: language, default mode, dorsal attention, visual, ventral attention, motor and fronto-parietal networks. An excel file (*Perceptron_ROI_list.xlsx*) can be found in the documentation with all the information related to seed positioning.

### Seed-to-seed measurements

Seed-to-seed data analysis provides measurements of functional connectivity between all the different pairs of seeds demonstrating a more global perspective about networks compared to seed-based functional connectivity. NeoRS performs Pearson correlation between the different ROIs to create a correlation matrix.

# 3. Results

## Image Registration

Image registration results of the T2-weighted and BOLD images to the template for a representative subject are demonstrated in figure 4, an example of a user checkpoint. Yellow lines represent the segmented cerebrospinal fluid, and is added as an overlay to the T2-weighted images,

BOLD and template. After visual inspection, a correct alignment within the template for both registrations was observed for each test participant.

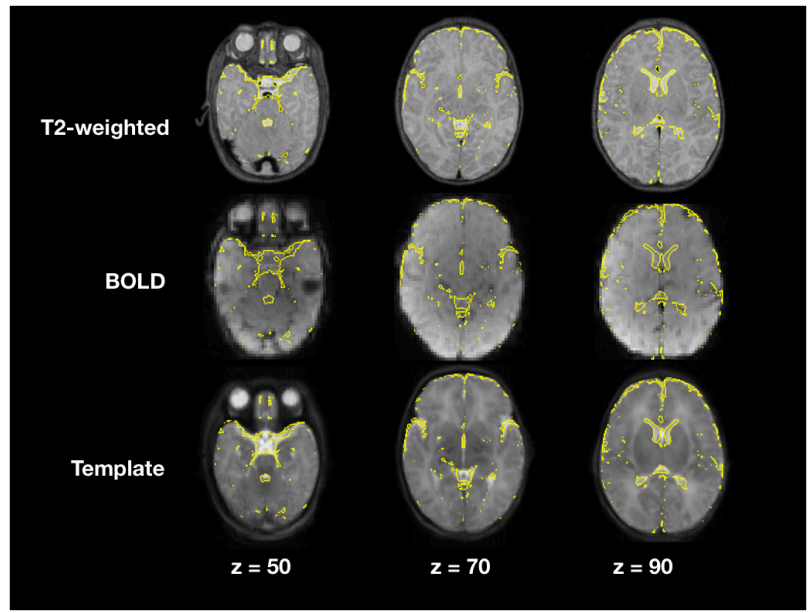

**Figure 4** T2-weighted and BOLD image registration to stereotaxic space. Gray scale images represent the T2-weighted, average BOLD and template images; the yellow lines correspond to the cerebrospinal fluid contours obtained from the T2-weighted image segmentations.

When comparing single step registration versus a 2-step registration approach for rsfMRI there were no discernible differences between both registrations and final functional connectivity results presented the same correlation strengths and topology. The main difference between the two approaches was computational times, which were higher for the 2-step registration method.

## Skull stripping

Figure 5 illustrates the default skull stripping segmentations versus NeoRS adapted parameters for neonates. With the default settings we observed skull stripping was failing for some of the subjects with different brain sizes, in contrast, when using NeoRS parameters, skull stripping remained robust for all the processed subjects.

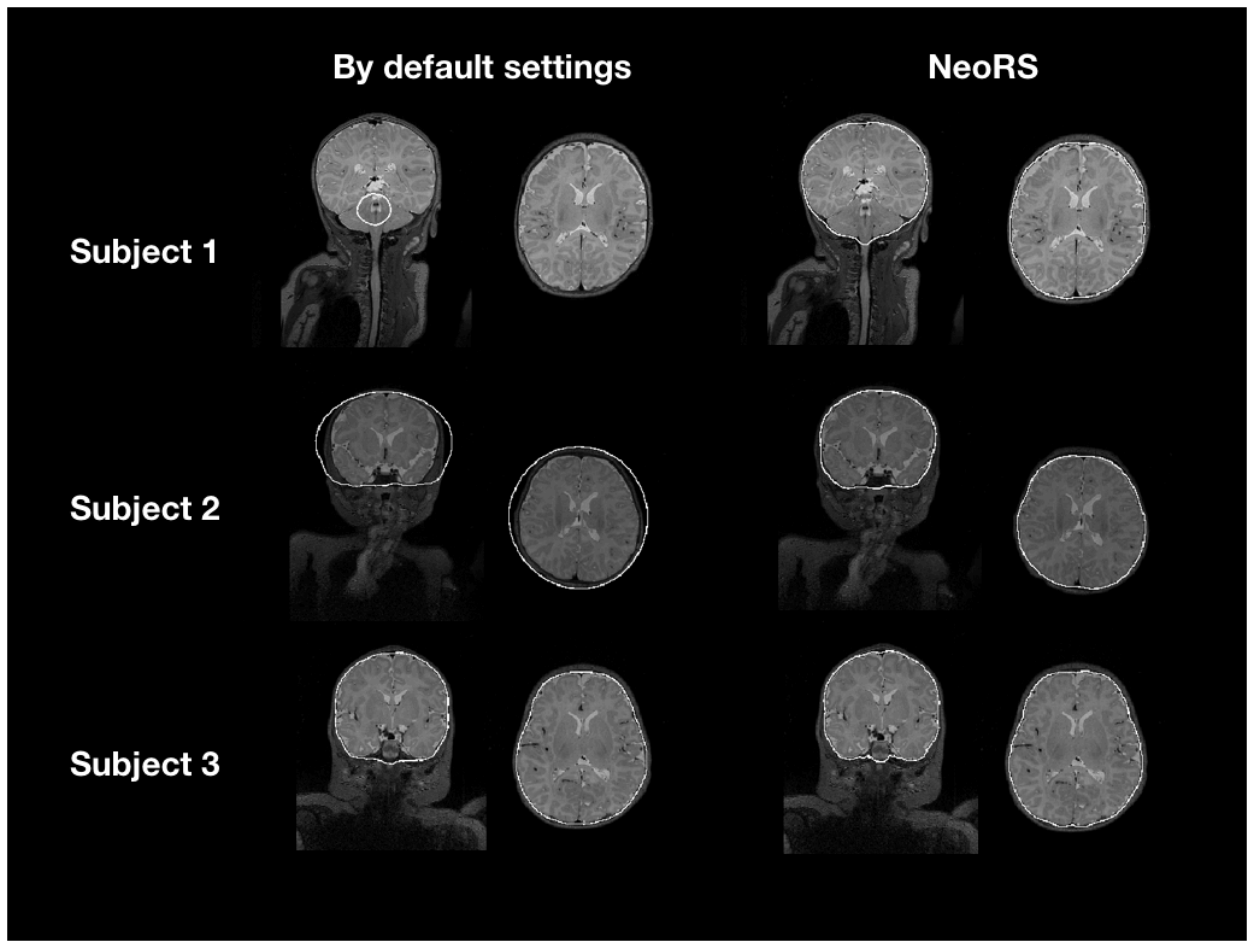

**Figure 5**. Skull stripping parameters comparing default bet2 settings in neonates and NeoRS settings optimized for neonates.

## Segmentation and Mask Creation

Figure 6 displays the 1 mm isotropic binary masks created by NeoRS from Mantis tissue probability maps. The output contains three different binary files corresponding to white matter, cerebrospinal fluid and gray matter. Figure 7 demonstrates the 3 mm isotropic masks for the regression of confounds process.

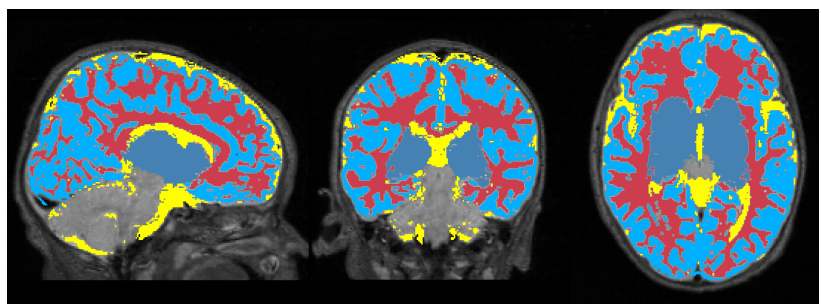

**Figure 6**. 1mm isotropic masks created from the tissue probability maps obtained with Mantis. White matter (red), csf (yellow), gray matter (blue).

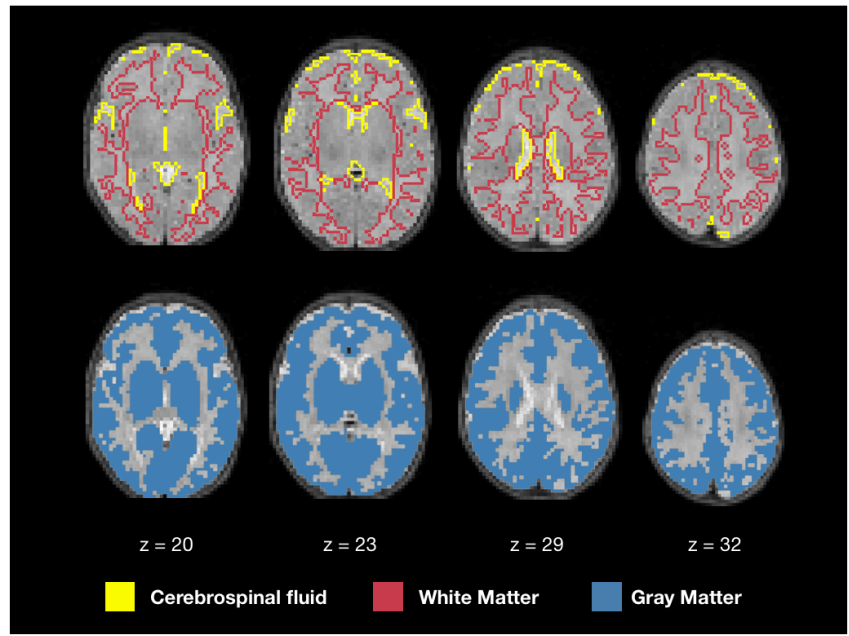

**Figure 7.** White matter, cerebrospinal fluid and gray matter masks for regression of confounds.

## Functional Distortion Correction

Functional susceptibility-induced magnetic field inhomogeneity correction for a representative subject is shown in figure 8. GRE EPI images, independent of brain size are distorted in the phase encoding direction whether they are acquired AP or PA but present those distortions in both areas of the brain. After susceptibility-induced magnetic field inhomogeneity distortion correction, the two acquisitions (AP and PA) present a similar morphology and a more accurate brain shape with respect to the undistorted T2-weighted image.

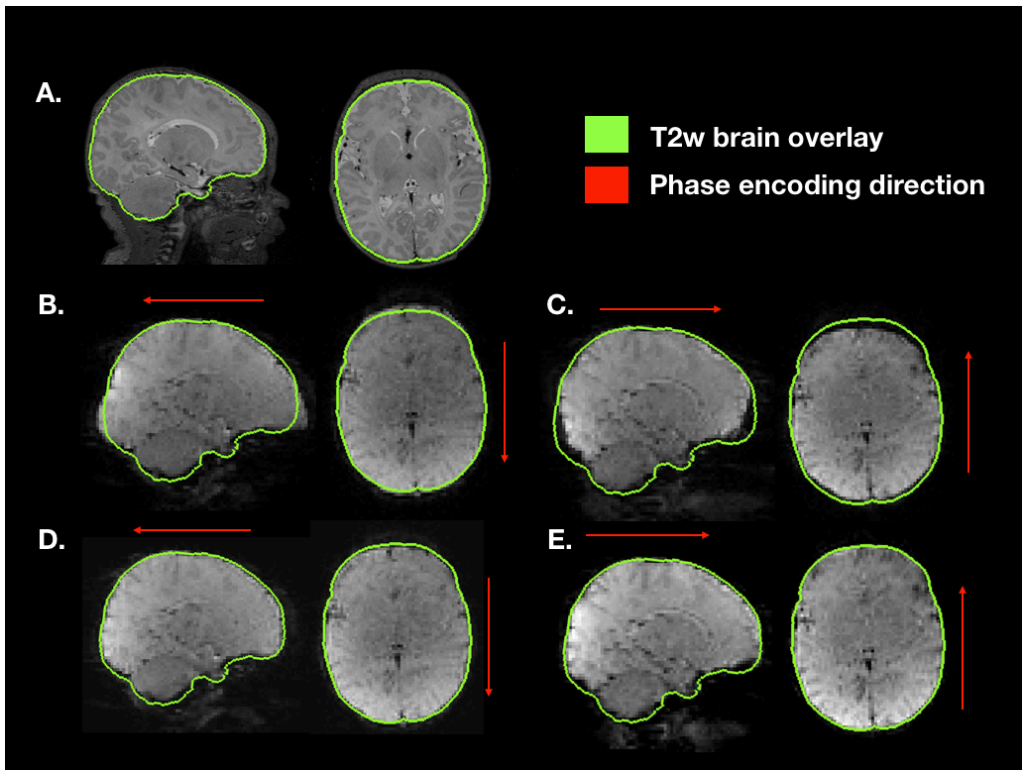

**Figure 8**. Susceptibility-induced magnetic field inhomogeneity causing geometric distortions along the phase encoding direction. A. Original T2-weighted image without distortion, shown as reference; B. Original GRE-EPI acquired in anterior-posterior phase encoding direction (AP); C. Original GRE-EPI acquired in posterior-anterior phase encoding direction (PA); D. Corrected GRE-EPI AP; E. Corrected GRE-EPI PA.

## Head Motion

After functional cross-realignment, an output graph is provided by NeoRS containing information concerning rotations and translations applied to cross-realign for each volume of the rsfMRI, as well as the computed framewise displacement [figure 9].

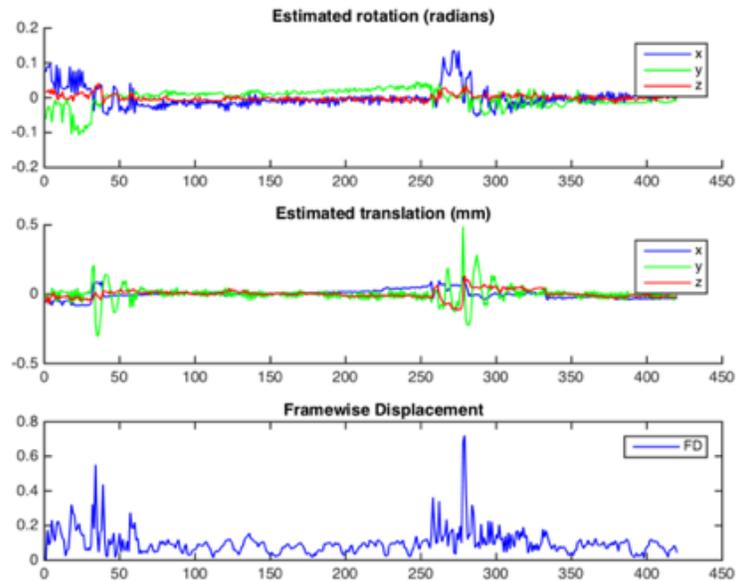

**Figure 9**. Example of head motion plots from a single subject. Plots are generated for each bold run and contain three different graphs per run: estimated rotation in radians; estimated translation in millimeters; Framewise displacement in millimeters.

Figure 10 is an example of a single subject with 2 different rsfMRI acquisitions with different amounts of motion. In the seed-based functional connectivity results of the motor network, the correlation differences between an acquisition with an average framewise displacement higher than 0.25 mm (run 1) and an acquisition with an average framewise displacement lower than 0.25 mm (run 2). High motion acquisition presented increased amounts of noise and the network topology was difficult to identify.

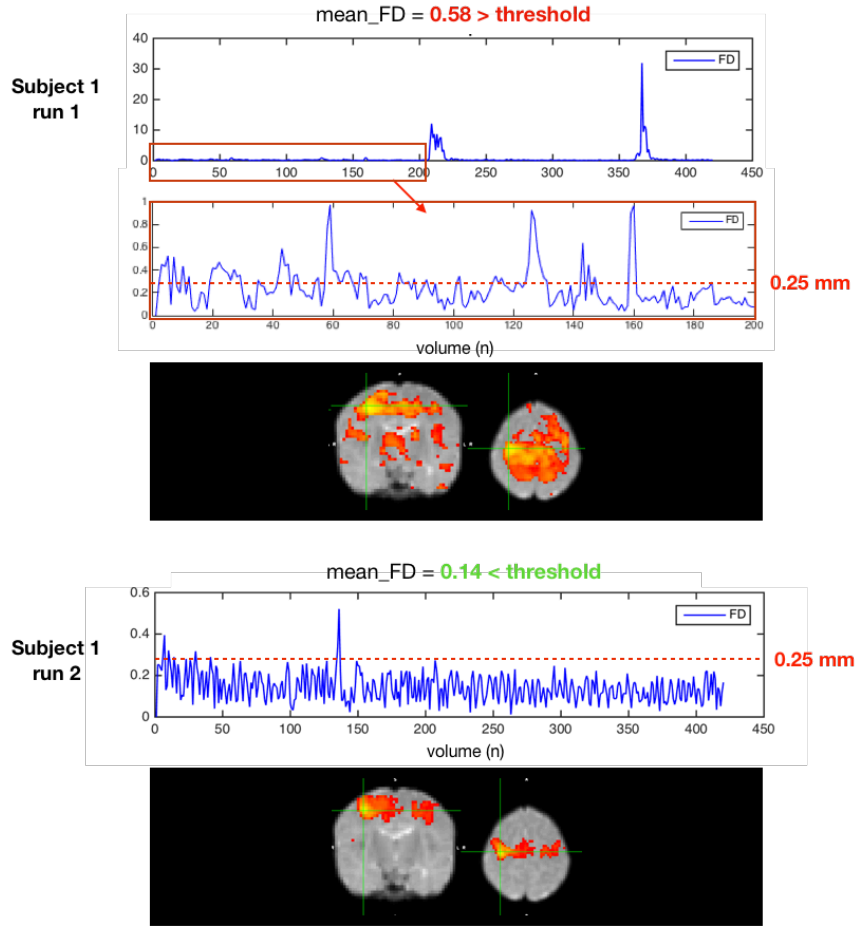

**Figure 10**. Two acquisitions of a high motion subject, run 1 excluded for having an average FD ≥0.25 mm, run 2 kept with an average FD < 0.25 mm.

## Resting State Networks - Seed-Based Correlations

Figure 11 illustrates, seven of the most common resting state networks after NeoRS processing employing an SBC approach.

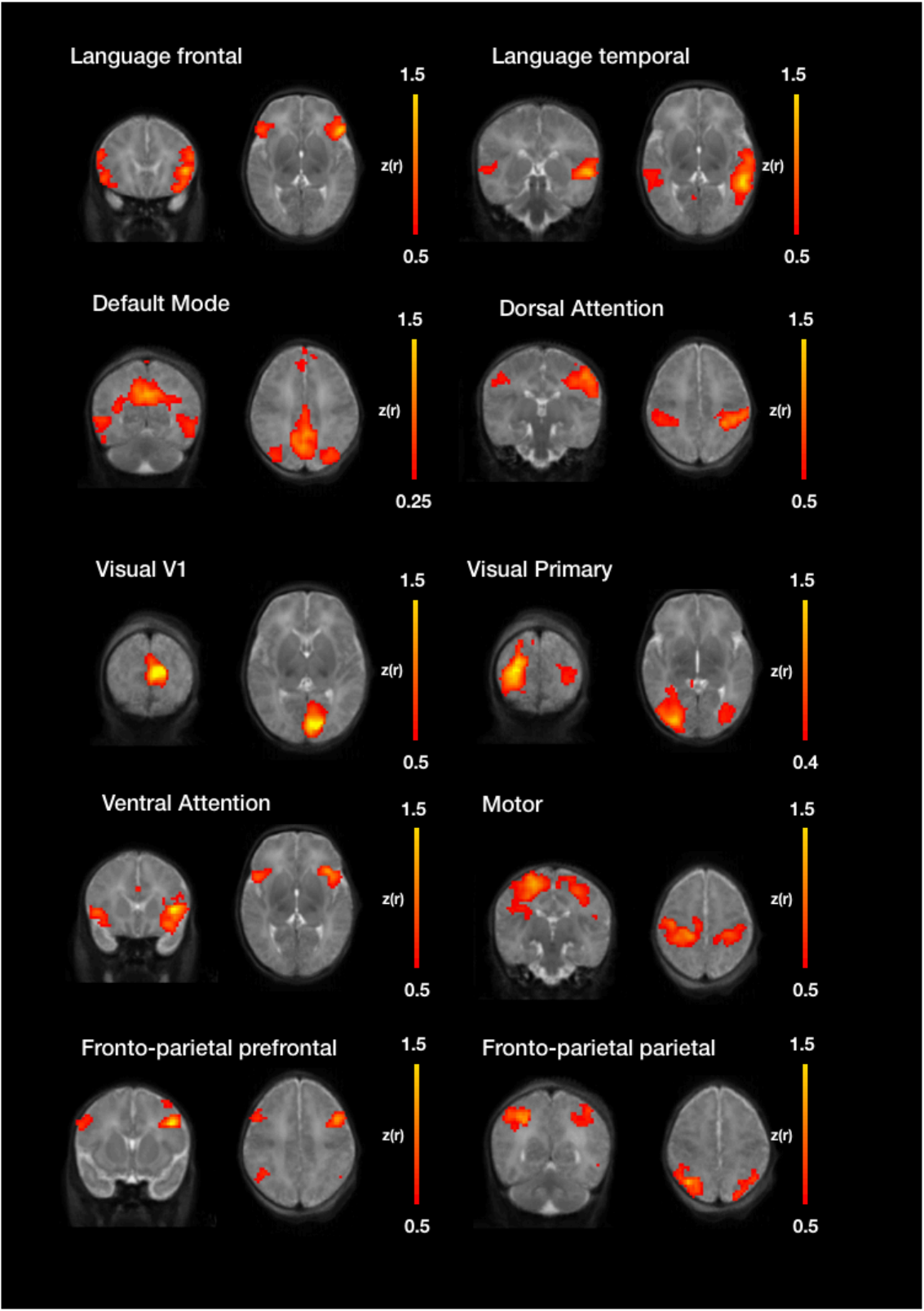

**Figure 11.** Example resting state networks obtained by seed-based functional connectivity after image processing with NeoRS.

Figure 12 is a single subject example of the 31 seeds included in NeoRS

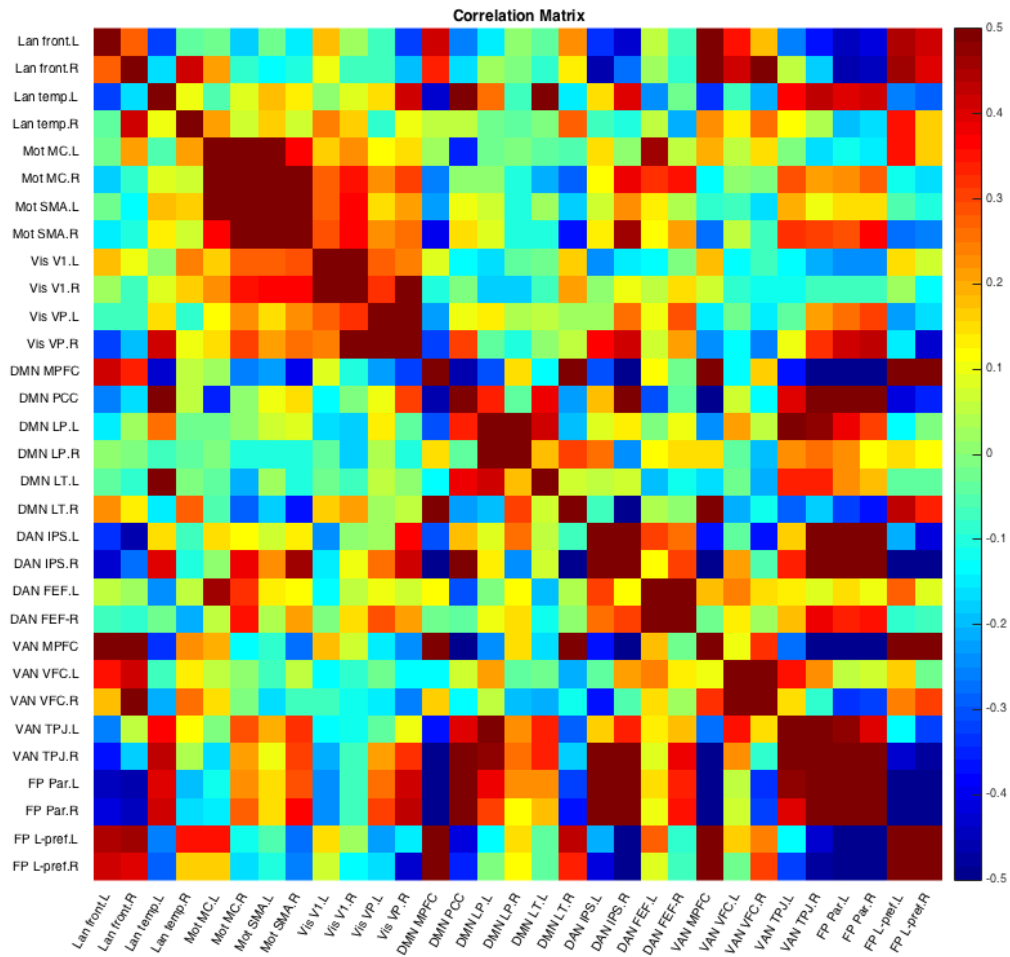

**Figure 12** Representative subject resting state network example seed-to-seed functional connectivity correlations.

## 4. Discussion

NeoRS is an open-source image processing pipeline dedicated to neonatal rsfMRI. It includes seed-based and seed-to-seed 1st level analysis. NeoRS has neonatal brain templates for term in 1mm and 3mm Talairach space, as well as a set of 31 seed regions defining seven common resting state networks. NeoRS relies on the open-source neuroimaging pipelines: SPM, FSL and AFNI and encompasses robust methods to segment, register and denoise neonatal rsfMRI data. Each of the various processing steps were evaluated separately. Output was carefully inspected to ensure the best quality products, by optimizing skull stripping, image registration, head motion and denoising related results. Additionally, functional connectivity of the motor, visual, default mode, language, dorsal attention, ventral attention and fronto-parietal networks using seed-based functional connectivity analysis were demonstrated.

Image registration to the atlas was meticulously inspected for every subject and no significant misalignments were observed for T2-weighted or rsfMRI images for affine registration. We compared single step to two-step registration for accuracy and computational times. The single step registration was chosen, as this process required less computational time and demonstrated no substantial variation when compared to the two-step counterpart.

In contrast with some of the most common adult pipelines, image registration is implemented prior to skull stripping, as alignment quality results were identical. Further, performing image registration preceding skull stripping produced vastly robust skull stripping results across varying brain sizes. This was not the case employing traditional skull stripping before image registration. Skull stripping of the T2-weighted images was found to be a crucial step when working with neonates because poor stripping lead to misclassification of segmented data and ultimately unreliable representation of RSN because of misregistrations. Additionally, image registration prior to skull stripping facilitated brain extraction without any user intervention. Furthermore, if the brain was previously aligned to an atlas, *bet2* was implemented on the subject specific atlas aligned data instead of applying an atlas brain mask to avoid subtle geometric inaccuracies. This procedure is critical and needs to be properly assessed. For this reason, an output for the skull stripping is provided which contains the image of the non-skull stripped mask with an overlay of the contour of the skull stripped brain.

After skull stripping, tissue segmentation is performed using *Mantis*, without the need of any further intervention. Mantis is integrated into the NeoRS pipeline as it provided robust results for disparate brain sizes using only T2-weighted images. Contrary to the adult brain that uses T1-weighted images for brain segmentation, it is fundamental in neonates to provide the pipeline with T2-weighted images as the water/cholesterol ratio is reversed with respect to adults due to lack of myelination. Neonatal T2-weighted images present a better contrast between brain structures [44]. While Mantis needs to be installed to use the NeoRS pipeline, no additional setup steps are required as it is fully assimilated in NeoRS. The results showed the binary masks created from *Mantis* tissue probability maps are perfectly aligned with their corresponding structures for various subjects without any manual intervention.

It is well known the GRE-EPI sequence for rsfMRI is prone to susceptibility-induced magnetic field inhomogeneity [45]. The artifact primarily appears close to the extrema portions of the brain in the phase encoding direction [46] and needs to be properly corrected. While several methods have been successfully used in these settings [47, 48], NeoRS operates the standard *topup/applytopup* method. Two reversed phase encoding direction images are involved to correct for the deformations. [49] The method is simple to implement in the acquisition protocol, provides high quality results and acquisition sequences require very short duration times.

Slice timing correction remains a controversial step when a very short repetition time (TR) is deployed [50]. However, as it has been shown it can significantly improve z-scores [50] and as the aim was to make NeoRS work with the maximum number of datasets, it is included as an option. Slice timing correction can be deactivated for multi-band sequences with very low TR, as in this kind of low TR sequence, all slices in each volume are acquired very closer together [23].

After data preprocessing, confounding signals and motion effects are removed. First, the framewise displacement threshold is defined as 0.25 mm, as performed by Smyser et al. on their neonatal study and shown to provide accurate results [9]. Volumes with FD higher than 0.25 mm were removed from the time-series as described per Power et al.[38]. Motion censoring was performed prior to filtering to prevent spikes from passing through band-pass filtering, as this could introduce artifacts such as Gibb's ringing and or skew correlation coefficients. Furthermore, extremely high motion acquisitions shouldn't be taken into account as they can lead to inflated results. To do so, different metrics can be adopted, such as maximum framewise displacement, minimum number of low motion volumes or average framewise displacement. In NeoRS, acquisitions with average FD higher than 0.25 mm were introducing augmented correlations related to motion and improper denoising. Those acquisitions were completely removed. Therefore, the average framewise displacement was employed as the metric for exclusion.

To correct for nuisance variables NeoRS implements a traditional denoising strategy that performs global signal, white matter and cerebrospinal fluid signal regression, motion parameter regression and a band-pass filter. This simple approach provides robust results [41], and doesn't require manual intervention or large datasets for denoising purposes. In contrast to the aforementioned independent components denoising techniques, [26, 27] such as the one employed in the dHCP pipeline.

Finally, NeoRS incorporates seed-based functional connectivity analysis tools to assist the user in assessing initial results. Seed-based results across subjects showed patterns very similar to those observed in the literature for all the analyzed resting state networks [11, 51] [figure 11]. Further data analysis can be carried out on the fully processed data, final_BOLD.nii, if desired.

## Limitations and Future Directions

### Limitations

NeoRS was fully vetted on the BCP, and CHU (results not shown) cohorts for neonates less than 9 weeks old. Expanding NeoRS to a larger cohort in both size and neonatal age variation, i.e. preterm, would only further demonstrate its novelty and application, but the current number of subjects is a limitation. Obviously, this expansion would also necessitate additional age specific atlases. Additionally, the pipeline requires a usable T2-weighted structural image and would benefit from the adaptability of the option of using a T1-weighted image, especially for younger neonates older than 9 weeks where tissue boundaries may begin to vary. NeoRS has been developed based on a traditional denoising strategy which includes band-pass filtering. This is a limitation on data acquired with very low TR, such as the dHCP data, as the degrees of freedom might be highly reduced. In the Fourier domain, the maximum sample rate corresponds to the frequency of Nyquist, $f_{max} = 1/2TR$, and the frequency spacing $\Delta f=1/t_{max}$, where $t_{max}$ is the total scan duration. If there are two data sets with the same total acquisition time, but different TR, the one with the lower TR will lose more degrees of freedom when filtering [52]. Another limitation of the pipeline is brains with significant malformations or injury. While a high range of brain sizes is accepted in the pipeline, anomalies such as neonatal hydrocephalus may require special attention, as part of the automated process could fail, such as cortical extraction. To overcome this limitation the user should perform rigorous individual image quality control of those brains and may adjust parameters, such as the head radius.

## 5. Conclusion

NeoRS [https://github.com/venguix/NeoRS] is an open-source, straightforward to use rsfMRI data processing pipeline for the neonatal brain which relies on the open-source neuroimaging pipelines FSL, AFNI and SPM. NeoRS works with neuroimaging nifti format, BIDS folder structures and has been developed to work with different MRI vendors and diverse acquisition parameters with minimal user implication. After image processing with NeoRS, we observed resting state networks were in agreement with previously published studies at term age. Each processing step is easy to inspect to ensure consistent results through quality control checkpoint figures.

An open-source, rudimentary to use pipeline for neonatal resting state image processing will allow the community to process their data immediately after scanning sessions implementing a simple computational infrastructure. The democratization of rsfMRI processing will allow a higher number of centers to collaborate and process their datasets and optimistically bring the clinical biomarker application one step closer.

## Acknowledgements

We would like to thank the Washington University – School of Medicine for sharing their templates for this project, and Jed Elison for letting us use the neonatal data from the Baby Connectome Project. We would also like to thank the Québec Bio-Imaging Network for supporting VE with studentship and funding for travels.

## Authors contributions

VE developed NeoRS, processed and analyzed the data, and wrote the manuscript. JK provided guidance and helped comparing results with a non open-source pipeline. DL contributed writing and reviewing the first draft and provided guidance, JCA provided guidance and reviewed the manuscript, GA supervised the project, contributed writing and reviewing all drafts and provided guidance. All authors contributed to manuscript revision, read, and approved the submitted version.